\documentclass[12pt]{iopart}

\usepackage{iopams}
\usepackage{graphicx}
\begin{document}
\title{Electromagnetically induced transparency in an inverted Y-type four-level system}
\author{Jianbing Qi}
\address{Department of Physics and Astronomy, Penn State University, Berks Campus, Tulpehocken Road, P.O. Box 7009, Reading, PA 19610}
\eads{jxq10@psu.edu}

\begin{abstract}
The interaction of a weak probe laser with an inverted-Y type four-level atomic system driven by two additional coherent fields is investigated theoretically. Under the influence of the coherent coupling fields, the steady-state linear susceptibility of the probe laser shows that the system can have single or double electromagnetically induced transparency windows depending on the amplitude and the detuning of the coupling lasers. The corresponding index of refraction associated with the group velocity of the probe laser can be controlled at both transparency windows by the coupling fields. The propagation of the probe field can be switched from superluminal near the resonance to subluminal on resonance within the single transparency window when two coupling lasers are on resonance. This provides a potential application in quantum information processing. We propose an atomic $^{87}Rb$ system for experimental observation.
\end{abstract}
\date{\today}
\pacs{42.50.Gy, 42.25.Kb}


\maketitle
\section{Introduction}
In recent years, substantial attention has been paid to the study of coherence effects in atomic and molecular systems~\cite{Nature.397.594.1999,PRA.60.3225.1999,PRA.63.023818.2001,JMO.43.471.1998,PRL.98.153601.2007,PRL.82.5229.1999, PRA.76.053420.2007,PRA.78.013830.2008,PRL.88.173003.2002}. The interaction of the coherent light with multilevel atomic ensembles results in many striking quantum phenomena. Three-level atomic systems, such as Lambda, Vee, and cascade schemes are the most widely used level schemes for study~\cite{OptCom.156.133.1998,PRA.71.053806.2005,PRA.71.063805.2005,PRA.75.053802.2007}. Among them, the electromagnetically induced transparency(EIT), which is based on the phenomenon of coherent population trapping~\cite{EArimondo.1996}, has attracted considerable attention~\cite{PRA.46.R29.1992,PRA.51.576.1995,PhysToday50.736.1997,Nature.413.273.2001,PRA.66.015802.2002,PRA.67.013811.2003,PRA.76.041802R.2007,PRA.73.043810.2006}. Some extraordinary effects associated with EIT have been studied theoretically, and observed experimentally, including ultraslow pulse propagation of light~\cite{Nature.397.594.1999,Ephys.35.2004}, light storage in an atomic vapor~\cite{Ephys.35.2004,PRA.78.023801.2008}, superluminal light propagation~\cite{PRA.71.053806.2005,Nature.406.277.2000,PRA.63.053806.2001,sci.301.200.2003,nature.425.695.2003}, coherent control of the optical information processing with BEC~\cite{Nature.445.623.2007}. An EIT system that results from a quantum interference effect can dramatically reduce the group velocity of a propagating probe laser with greatly reduced or even vanishing absorption of the probe laser. The essential physical mechanism is that the internal structure of atoms can be modified by the interaction with both the coupling(or $"$control$"$) field and the probe field. Therefore the interaction of the probe field with the atoms can be manipulated by the coupling field~\cite{JPhB.35.R31.2002}. A properly chosen and prepared atomic system is essential for a successful experimental observation. Alkali atoms have served as test species for these effects due to the availability of laser wavelength and spectroscopy data. Multilevel rubidium atomic systems provide an excellent test ground and a starting point for extending the control dimensions with inexpensive available laser frequencies for its atomic energy structures.

In this paper, we investigate the response of a probe laser in an inverted Y-type four-level system driven by two additional coherent fields. This scheme has been used in the study of the Autler-Townes effect in a sodium dimer~\cite{JCP124.084308.2006}, and of two-photon fluorescence suppression in an ultracold rubidium atom~\cite{PRA.64.043807.2001}, respectively. Here, we study the absorption and dispersion of a weak probe laser using probability amplitude and equivalent density matrix methods to obtain the linear susceptibility of the probe laser. We show that the system exhibits two electromagnetically induced transparency windows for the probe field. The transparency windows can be controlled by the amplitudes and frequency detunings of the coupling fields. The index of refraction associated with the group velocity of the probe laser can be very different at the two transparency windows and can be controlled by the two coupling fields. We propose an atomic $^{87}Rb$ system for experimental observation of this phenomenon.

\section{Equations of Motion}
We consider an inverted-Y type four-level system interacts with three lasers, L1, L2 and L3 as shown in figure 1(a). Two ground states, $|1\rangle$ and $|2\rangle$, are coupled by laser L1 and L2 to a common excited state $|3\rangle$, and the excited state $|3\rangle$ is coupled by laser L3 to an upper excited state $|4\rangle$, respectively. laser L1 is a weak probe laser, L2 and L3 are two coupling (or $"$control$"$) lasers. The corresponding dressed-state($|+\rangle$, $|0\rangle$, and $|-\rangle$)diagram of two coupling lasers interacting with level $|2\rangle$, $|3\rangle$ and $|4\rangle$ is shown in figure 1(b). All the transitions are electric dipole allowed. The Hamiltonian of the system is given by
\begin{equation}
    H = H_{0}+H_I\label{eq1},
\end{equation}
where
\begin{equation}
H_{0} = \sum_{i=1}^{4} {\hbar\omega_{i}|i \rangle\langle i|}\label{eq2}
\end{equation}
is the atomic Hamiltonian, $\hbar\omega_{i}$ is the energy of the isolated atom in state $|i\rangle$, and $H_I$ is the dipole interaction Hamiltonian, which is given by

\begin{equation}
H_I=\sum_{i\neq{j}} \langle i|(-\vec{\mu}\cdot\vec{E})|j\rangle
    =-\sum_{i\neq{j}}\mu_{ij}E_{ij}\label{eq3},
\end{equation}
where $\mu_{ij}$ is the electric dipole moment for $|i\rangle\leftrightarrow|j\rangle$ transition, and $E_{ij}$ is the corresponding coupling laser field. In the rotating-wave-approximation, the interaction Hamiltonian can be written as:
\begin{eqnarray}
H_I&=&-\frac{\hbar}{2}(\Omega_{1}e^{-i\nu_{1}t}|3\rangle\langle{1}|
    +\Omega_{2}e^{-i\nu_{2}t}|3\rangle\langle{2}|\nonumber\\
    &&+\Omega_{3}e^{-i\nu_{3}t}|4\rangle\langle3|)+h.c.\label{eq4},
\end{eqnarray}
where $\nu_{i}$ is the laser frequency and  $\Omega_{i}=\mu_{ij}E_{ij}/\hbar$ is the corresponding Rabi frequency which is assumed positive in our calculation.
\begin{figure}
  \includegraphics{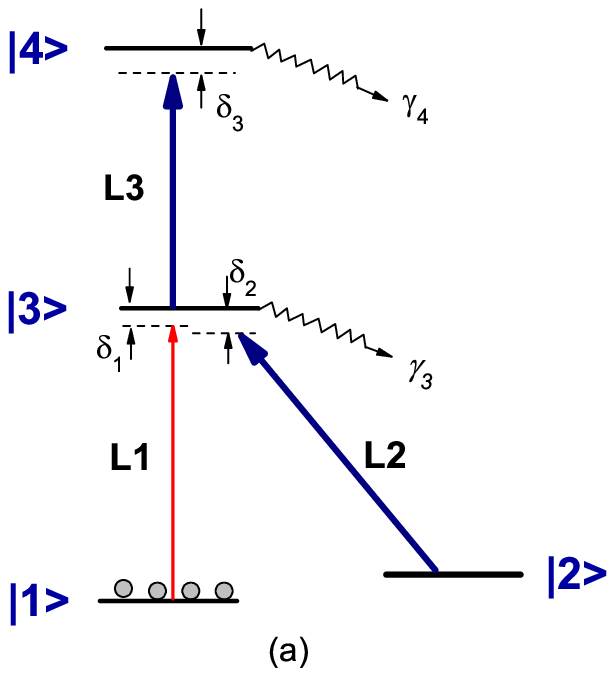}
  \includegraphics{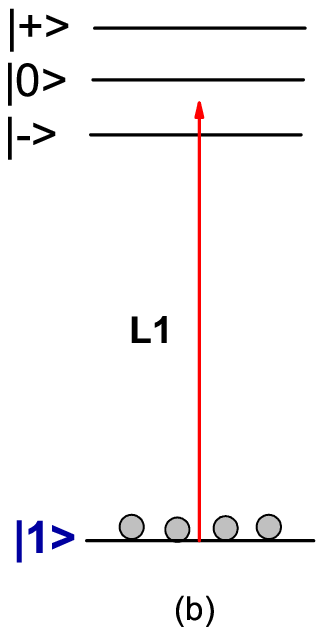}
\caption{(Color online)(a) Energy level scheme for an inverted Y-type four-level atom. (b) Corresponding dressed-state diagram of laser L2 and L3 interacting with $|2\rangle$, $|3\rangle$, and $|4\rangle$.}
\end{figure}
The Hamiltonian of the system in the interaction representation can be written as
\begin{eqnarray}
H_{int}&=&-\frac{\hbar}{2}(\Omega_{1}e^{-i\delta_{1}t}|3\rangle\langle{1}|
    +\Omega_{2}e^{-i\delta_{2}t}|3\rangle\langle{2}|\nonumber\\
    &&+\Omega_{3}e^{-i\delta_{3}t}|4\rangle\langle3|)+h.c.\label{eq5},
\end{eqnarray}
where $\delta_{1}=\nu_{1}-\omega_{31}$, $\delta_{2}=\nu_{2}-\omega_{32}$, and
$\delta_{3}=\nu_{3}-\omega_{43}$ are the frequency detunings of the probe laser L1, coupling lasers L2, and L3, respectively. $\omega_{ij}=\omega_i-\omega_j$ is the $|i\rangle\leftrightarrow|j\rangle$ resonance transition frequency. We assume $\omega_{1}=0$ for simplicity and the energy of all other states are measured relative to state $|1\rangle$.
\subsection{Probability Amplitude Approach}
The atomic wave-function of the system in the interaction picture at any time t can be expanded in terms of bare-state eigenvectors as
\begin{eqnarray}
|\Psi_{int}(t)\rangle=a_{1}(t)|1\rangle
                    + a_{2}(t)|2\rangle
                    + a_{3}(t)|3\rangle
                    + a_{4}(t)|4\rangle\label{eq6},
\end{eqnarray}
where $a_{i}(t)$ is the time-dependent probability amplitude of the atomic state $|i\rangle$. The Schr\"{o}dinger equation in the interaction picture reads as
\begin{eqnarray}
\frac{\partial{|\Psi_{int}(t)\rangle}}{\partial{t}}=-\frac{i}{\hbar}H_{int}|\Psi_{int}(t)\rangle\label{eq7}
\end{eqnarray}
By introducing the wave-function $\Psi_{int}$, and the interaction Hamiltonian $H_{int}$ of equation (5) into the Schr\"{o}dinger equation, and after making some rotating transformations, we obtain the equations for the evolution of probability amplitudes of the wave function as follows:
\numparts
\begin{eqnarray}
\dot{a}_{1}(t) = i\frac{\Omega_{1}}{2}a_{3}(t)\label{eq(8a)}
\end{eqnarray}
\begin{eqnarray}
\dot{a}_{2}(t)=i(\delta_{1}-\delta_{2})a_{2}(t)
                +i\frac{\Omega_{2}}{2}a_{3}(t)\label{eq8b}
\end{eqnarray}
\begin{eqnarray}
\dot{a}_{3}(t) = i\delta_{1}a_{3}(t)
                +i\frac{\Omega_{1}}{2}a_{1}(t)
                +i\frac{\Omega_{2}}{2}a_{2}(t)
                +i\frac{\Omega_{3}}{2}a_{4}(t)\label{eq8c}
\end{eqnarray}
\begin{eqnarray}
\dot{a}_{4}(t) = i(\delta_{1}+\delta_{3})a_{4}(t)
                +i\frac{\Omega_{3}}{2}a_{3}(t)\label{eq8d}
\end{eqnarray}
\endnumparts
We assume that the probe laser is weak and the population is initially in level $|1\rangle$, $a_1=1$. We solve the above equations to the first order in terms of the Rabi frequency $\Omega_{1}$ of the probe laser , and to all orders in $\Omega_{3}$ and $\Omega_{2}$ of the coupling lasers under the steady-state condition. From equation (8b)-(8d) we obtain
\numparts
\begin{eqnarray}
\dot{a}_{2}^{(1)}(t)-i(\delta_{1}-\delta_{2})a_{2}^{(1)}(t)
    =i\frac{\Omega_{2}}{2}a_{3}^{(1)}(t)\label{eq9a}
\end{eqnarray}
\begin{eqnarray}
\dot{a}_{3}^{(1)}(t)-i\delta_{1}a_{3}^{(1)}(t)
                =i\frac{\Omega_{1}}{2}
                +i\frac{\Omega_{2}}{2}a_{2}^{(1)}(t)
                +i\frac{\Omega_{3}}{2}a_{4}^{(1)}(t)\label{eq9b}
\end{eqnarray}
\begin{eqnarray}
\dot{a}_{4}^{(1)}(t)-i(\delta_{1}+\delta_{3})a_{4}^{(1)}(t)
    =i\frac{\Omega_{3}}{2}a_{3}^{(1)}(t)\label{eq9c}
\end{eqnarray}
\endnumparts
The steady-state solution of $a_3^{(1)}$ is given by
\begin{eqnarray}
a_3^{(1)}=\frac{\Omega_{1}}{2}
        \left(-\delta_1
        +\frac{\frac{\Omega_{2}^2}{4}}{\delta_1-\delta_2}
        +\frac{\frac{\Omega_{3}^2}{4}}{\delta_1+\delta_3}
        \right)^{-1}\label{eq10}.
\end{eqnarray}
The susceptibility at the probe frequency is given by
\begin{eqnarray}
\chi=\frac{2N\mu_{13}a_{1}^{*}a_{3}}{\epsilon_{0}E_1}\nonumber\\
    =\frac{N|\mu_{13}|^2}{\epsilon_{0}\hbar}
       \left(-\delta_1
        +\frac{\frac{\Omega_{2}^2}{4}}{\delta_1-\delta_2}
        +\frac{\frac{\Omega_{3}^2}{4}}{\delta_1+\delta_3}
        \right)^{-1}\label{eq11},
\end{eqnarray}
where N is the atomic number density.
Now we include the effects of damping using a phenomenological description. Let $\gamma_3/2$ and $\gamma_4/2$ be the decay rates of the probability amplitude of levels $|3\rangle$ and $|4\rangle$, respectively. By inspecting equation (9b) and (9c) we can see that the effects of damping can be included by replacing $\delta_1$ by $\delta_1+i\gamma_3/2$ in equation (9b) and $\delta_1+\delta_3$ by $\delta_1+\delta_3+i\gamma_4/2$ in equation (9c). This results in the susceptibility as
\begin{eqnarray}
\chi=\frac{N|\mu_{13}|^2}{\epsilon_{0}\hbar}
      \left(-\delta_1-\frac{i\gamma_3}{2}
        +\frac{\Omega_{2}^2/4}{\delta_1-\delta_2}
        +\frac{\Omega_{3}^2/4}{\delta_1+\delta_3+i\gamma_4/2}\right)^{-1}\label{eq12}
\end{eqnarray}
\subsection{Density Matrix Equations}
We can also model this system by density matrix equations. The master equation of motion for the density operator in the interaction representation is given by
\label{density matrix eq.}
\begin{equation}
\frac{\partial\varrho}{\partial{t}}=-\frac{i}{\hbar}[H_{int},\varrho]
                                    +{(}\frac{\partial\varrho}{\partial{t}}{)}_{inc}\label{eq13},
\end{equation}
where the second term on the left hand-side  represents the damping due to spontaneous emission and other irreversible processes. In the rotating-wave-approximation it is straightforward to obtain the density matrix equations as follows:
\label{density matrix equation}
\begin{eqnarray}
\dot{\rho}_{44}=\frac{i\Omega_{3}}{2}(\rho_{34}
                -\rho_{43})
                -\gamma_{4}\rho_{44}\label{eq14}
\end{eqnarray}
\begin{eqnarray}
\dot{\rho}_{33}=&\frac{i\Omega_{1}}{2}(\rho_{13}-\rho_{31})
                +\frac{i\Omega_{3}}{2}(\rho_{43}-\rho_{34})
                +\frac{i\Omega_{2}}{2}(\rho_{23}-\rho_{32})\nonumber\\
                &+\gamma_4\rho_{44}-\gamma_3\rho_{33}\label{eq15}
\end{eqnarray}
\begin{eqnarray}
\dot{\rho}_{22}=\frac{i\Omega_{2}}{2}(\rho_{32}-\rho_{23})
                +W_{32}\rho_{33}-\gamma_2\rho_{22}\label{eq16}
\end{eqnarray}
\begin{eqnarray}
\rho_{11}+\rho_{22}+\rho_{33}+\rho_{44}=1\label{eq17}
\end{eqnarray}
\begin{eqnarray}
\dot{\rho}_{31}=(i\delta_1-\gamma_{31})\rho_{31}
                +\frac{i\Omega_{1}}{2}(\rho_{11}-\rho_{33})
                +\frac{i\Omega_{3}}{2}\rho_{41}
                +\frac{i\Omega_{2}}{2}\rho_{21}\label{eq18}
\end{eqnarray}
\begin{eqnarray}
\dot{\rho}_{41}=[i(\delta_1+\delta_3)-\gamma_{41}]\rho_{41}
                +\frac{i\Omega_{3}}{2}\rho_{31}
                -\frac{i\Omega_{1}}{2}\rho_{43}\label{eq19}
\end{eqnarray}
\begin{eqnarray}
\dot{\rho}_{12}=[i(\delta_2-\delta_1)-\gamma_{12}]\rho_{12}
                +\frac{i\Omega_{1}}{2}\rho_{32}
                -\frac{i\Omega_{2}}{2}\rho_{13}\label{eq20}
\end{eqnarray}
\begin{eqnarray}
\dot{\rho}_{32}=(i\delta_2-\gamma_{32})\rho_{32}
                +\frac{i\Omega_{1}}{2}\rho_{12}
                +\frac{i\Omega_{3}}{2}\rho_{42}
                +\frac{i\Omega_{2}}{2}(\rho_{22}-\rho_{33})\label{eq21}
\end{eqnarray}
\begin{eqnarray}
\dot{\rho}_{42}=[i(\delta_3+\delta_2)-\gamma_{42}]\rho_{42}
                +\frac{i\Omega_{3}}{2}\rho_{32}
                -\frac{i\Omega_{2}}{2}\rho_{43}\label{eq22}
\end{eqnarray}
\begin{eqnarray}
\dot{\rho}_{43}=(i\delta_3-\gamma_{43})\rho_{43}
                -\frac{i\Omega_{1}}{2}\rho_{41}
                +\frac{i\Omega_{3}}{2}(\rho_{33}-\rho_{44})
                -\frac{i\Omega_{2}}{2}\rho_{42}\label{eq23}
\end{eqnarray}
where $\gamma_i$ is the population decay rate of level $|i\rangle$, $W_{ij}$ is the branch decay rate from level $|i\rangle$ to $|j\rangle$, and $\gamma_{ij}$ represents the coherence decay rate which is given by $\gamma_{ij}=\gamma_{ji}=(\gamma_i+\gamma_j)/2+\gamma_{ij}^c$. $\gamma_{ij}^c$ the collision dephasing rate.

We assume that the population is initially in its ground state level $|1\rangle$, and the probe laser is weak so that $\rho_{11}{(0)}\approx1$ for all times. Again, we solve the equations to the first order of the Rabi frequency of the probe laser and to all orders of the coupling lasers in the steady-state condition. We are interested in the off-diagonal matrix element $\rho_{31}$ associated with the susceptibility of the probe field. By setting the derivatives to zero and keeping the first order term in $\Omega_1$ in equations (18)-(20) we obtain the steady-state solution of $\rho_{31}$ to the first order of $\Omega_{1}$
\begin{eqnarray}
\rho_{31}^{(1)}=\frac{\Omega_{1}\rho_{11}{(0)}}{2}
                \left(-\delta_1-i\gamma_{31}
                    +\frac{\Omega_{2}^2/4}{\delta_1-\delta_2+i\gamma_{12}}
                    +\frac{\Omega_{3}^2/4}{\delta_1+\delta_3+i\gamma_{41}}
                    \right)^{-1}\label{eq24}
\end{eqnarray}
Then the susceptibility reads
\begin{eqnarray}
\chi&=&\frac{N\mu_{13}\rho_{31}}{\epsilon_{0}E_{1}}\nonumber\\
    &=&\frac{N|\mu_{13}|^2}{\epsilon_{0}\hbar}
        \left(-\delta_1-i\gamma_{31}
        +\frac{\Omega_{2}^2/4}{\delta_1-\delta_2+i\gamma_{12}}
        +\frac{\Omega_{3}^2/4}{\delta_1+\delta_3+i\gamma_{41}}
        \right)^{-1}\label{25}.
\end{eqnarray}
The similarity of equation (25) to equation (12) is clear. If the decay rates of the ground states($|1\rangle$ and $|2\rangle$) and the collision dephasing rate are small compared to $\gamma_3$, and can be neglected, by setting $\gamma_1=\gamma_2=\gamma_{ij}^c=0$, equation (25) is identical to equation (12).
\section{Discussion and Numerical Results}
\subsection{Absorption Spectra}
It is well-known that the imaginary part of the susceptibility gives the absorption and the real part gives the dispersion of the probe field. The susceptibility $\chi$ in equation (12) can  be separated into the real($\chi^{'}$) and imaginary($\chi^{''}$) parts as $\chi=\chi^{'}+i\chi^{''}$. The explicit expressions for $\chi^{'}$ and $\chi^{''}$ are
\numparts
\begin{eqnarray}
\chi^{'}=\frac{N|\mu_{13}|^2}{\epsilon_{0}\hbar}
            \frac{(\delta_1-\delta_2)[A(\delta_1+\delta_3)+B\gamma_4/2]}
            {A^2+B^2}\label{26a}\\
\chi^{''}=\frac{N|\mu_{13}|^2}{\epsilon_{0}\hbar}
            \frac{(\delta_1-\delta_2)[A\gamma_4/2-B(\delta_1+\delta_3)]}
            {A^2+B^2}\label{26b},
\end{eqnarray}
\endnumparts
with
\begin{eqnarray}
A=&&(\delta_1+\delta_3)\frac{\Omega_{2}^2}{4}-(\delta_2-\delta_1)\frac{\Omega_{3}^2}{4}\nonumber\\
    &&+(\delta_2-\delta_1)[\delta_1(\delta_1+\delta_3)-\frac{\gamma_3\gamma_4}{4}]\nonumber\label{A}
\end{eqnarray}
\begin{eqnarray}
B=(\delta_2-\delta_1)[\delta_1\frac{\gamma_4}{2}+(\delta_1+\delta_3)\frac{\gamma_3}{2}]
    +\frac{\Omega_{2}^2\gamma_4}{8}\nonumber\label{B}.
\end{eqnarray}

In our following numerical calculations we assume the atoms are at ultracold temperatures and the Doppler effect can be neglected. An ultracold atomic sample can be obtained in a magnetic-optical-trap(MOT), such as an ultracold Rb atom trap. For example, $10^7\sim10^8$ atoms can be trapped in a $"$dark-spot$"$ MOT within a spherical cloud of a size of 1.0 mm in diameter,and the atomic number density N can be around $10^{10}\sim10^{11}/cm^3$~\cite{PRL.93.243005.2004,JPhB.39.s813.2006}. A transition dipole moment of $2.5\times10^{-29}Cm$ is used for $\mu_{13}$ in our calculations which is corresponding to the value of $^{87}Rb$ $D_1$ transition~\cite{http.steck.us.alklidata}. Combining these parameters, we obtain the coefficient $N|\mu_{13}|^2/(\epsilon_{0}\hbar)\approx1.28\times10^4/s$. For demonstration we assume in our model that the upper state decays much slower than the intermediate excited state $|3\rangle$, such as $\gamma_4=0.1\gamma_3$ in our calculation. We will discuss this with more details later in this paper.
\begin{figure}
    \includegraphics[width=8cm]{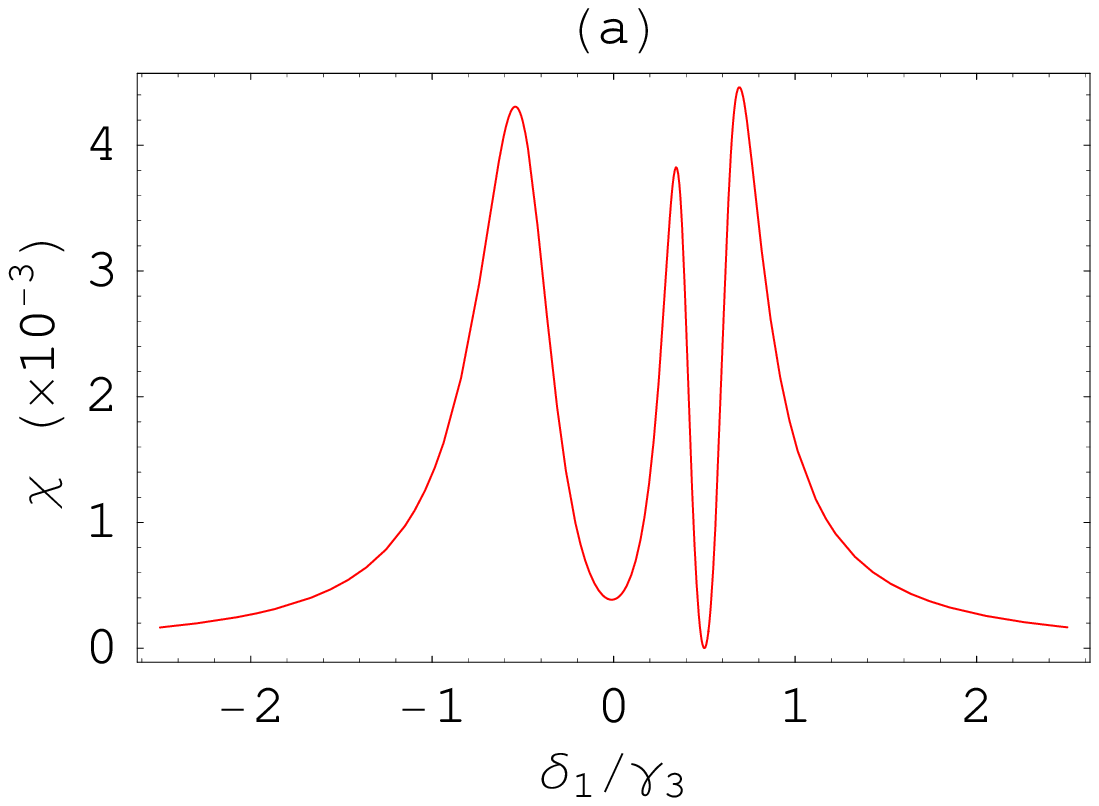}
    \includegraphics[width=8cm]{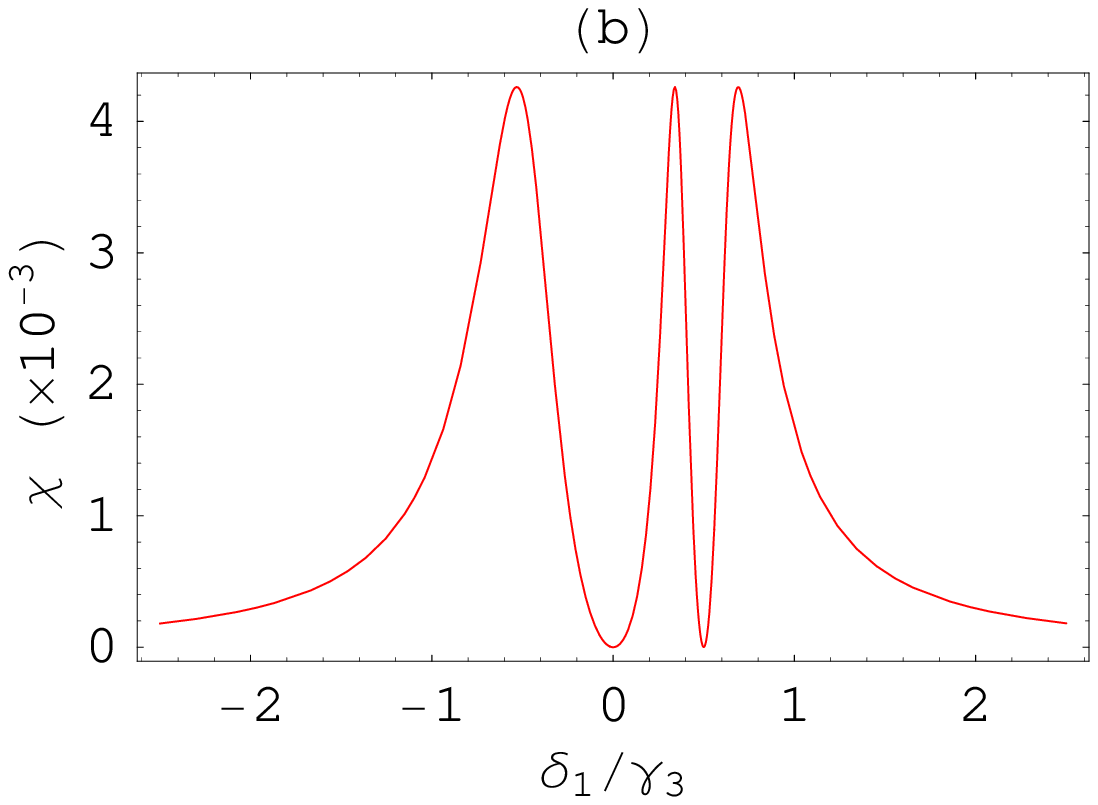}
    \includegraphics[width=8cm]{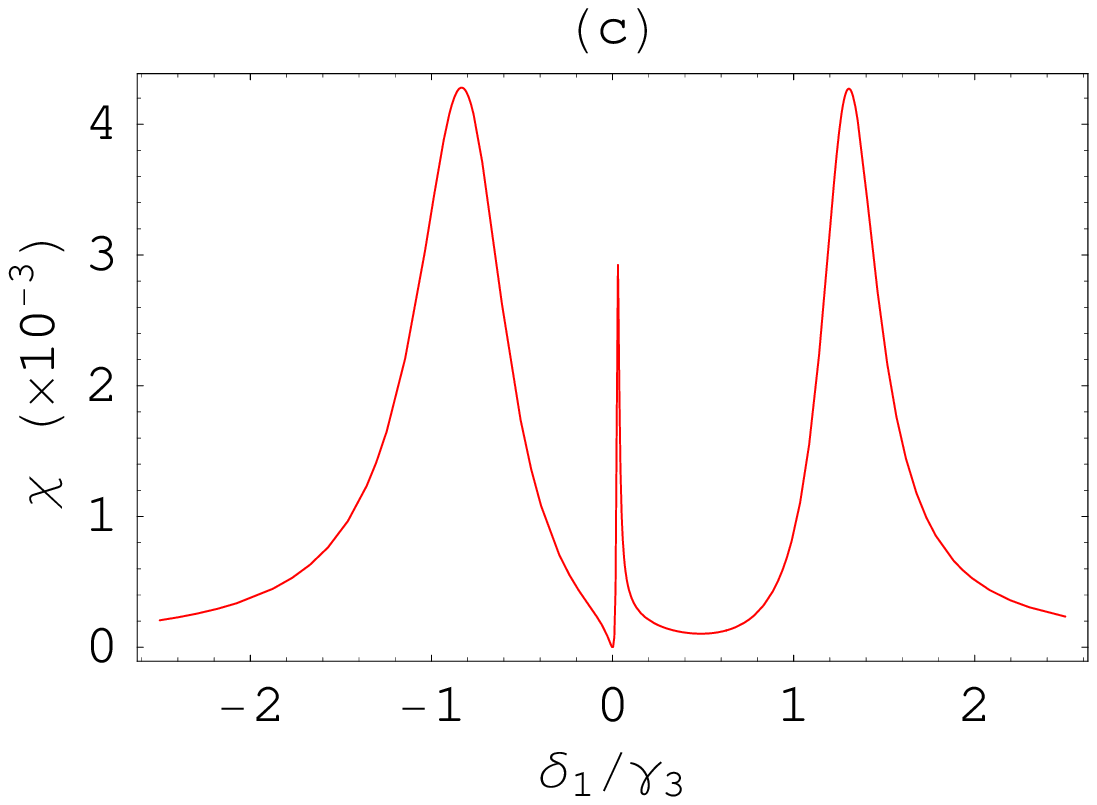}
    \includegraphics[width=8cm]{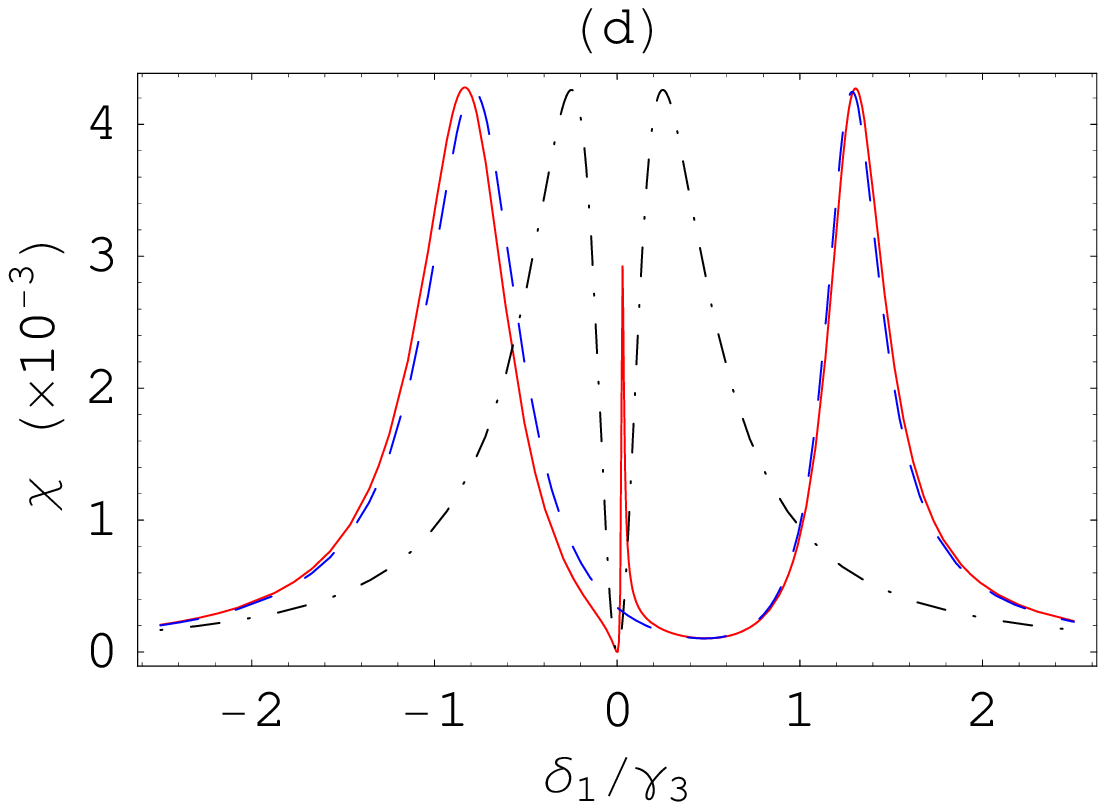}
\caption{(Color online) The imaginary part of the linear susceptibility of the probe laser as a function of the probe frequency detuning. There are two transparency windows at $\delta_1=\delta_2$, and $\delta_1=-\delta_3$, respectively. (a) $\delta_2=0.5\gamma_3$, $\Omega_{2}=0.5\gamma_3$, $\delta_3=0$, and $\Omega_{3}=\gamma_3$. (b) When $\gamma_4=0$ the probe laser at $\delta_1=-\delta_3$ is also completely transparent. Other parameters are the same as in (a). (c) $\delta_2=0$, $\Omega_{2}=0.5\gamma_3$, $\delta_3=-0.5\gamma_3$, and $\Omega_{3}=2\gamma_3$. (d) Solid line is the same as in (c); dashed lines are for the cascade scheme, $|1\rangle-|3\rangle-|4\rangle$: $\Omega_2=0$ and $\delta_3=-0.5\gamma_3$; dotdashed lines are for the $\Lambda$ scheme, $|1\rangle-|3\rangle-|2\rangle$: $\Omega_3=0$ and $\delta_2=0$.}
\end{figure}

By inspecting equation (26b), we see that the absorption spectrum of the probe laser can have two electromagnetically induced transparency windows as long as the two coupling lasers are neither on resonance simultaneously nor at a Raman detuning($\delta_2=-\delta_3$), that is, when the two coupling lasers have different frequency detunings, the absorption of the probe laser displays two minima at $\delta_1=\delta_2$ and $\delta_1=-\delta_3$, respectively, as demonstrated in figure 2(a)-(c). The absorption of the probe laser becomes zero(or the atomic system becomes completely transparent to the probe laser) as the probe laser frequency is detuned at $\delta_1=\delta_2$. The second minima is at $\delta_1=-\delta_3$ but the absorption is not completely zero at this detuning due to the decay of the upper excited state $|4\rangle$. If the upper state $|4\rangle$ does not decay, the absorption will also be zero at $\delta_1=-\delta_3$ as shown in figure 2(b). However, the absorption is significantly reduced at $\delta_1=-\delta_3$ even for a decaying upper excited state provided that the coupling laser L3 is strong with respect to the decay rate. If we choose an atomic system with a small decay rate $\gamma_4$ of the upper state, such as a metastable state, then the absorption of the probe laser can be reduced greatly by a relatively strong coupling laser L3 as illustrated in figure 2(c). Let us compare this scheme with the widely used three-level $\Lambda$ and cascade systems in the EIT study. Equation (26) can be used for a $\Lambda$ scheme $|1\rangle-|3\rangle-|2\rangle$ by setting $\Omega_3=0$, and for a cascade scheme $|1\rangle-|3\rangle-|4\rangle$ by setting $\Omega_2=0$, respectively. In both cases there is only one transparency window as illustrated in figure 2(d) by the dotdashed line and dashed line. The combination of both systems brings another dimension of control of EIT. As one can see from figure 2(d), an absorption line emerges within the transparency window of the $\Lambda$ scheme by introducing laser L3, and the absorption linewidth can be subnatural.
\begin{figure}
    \includegraphics[width=8cm]{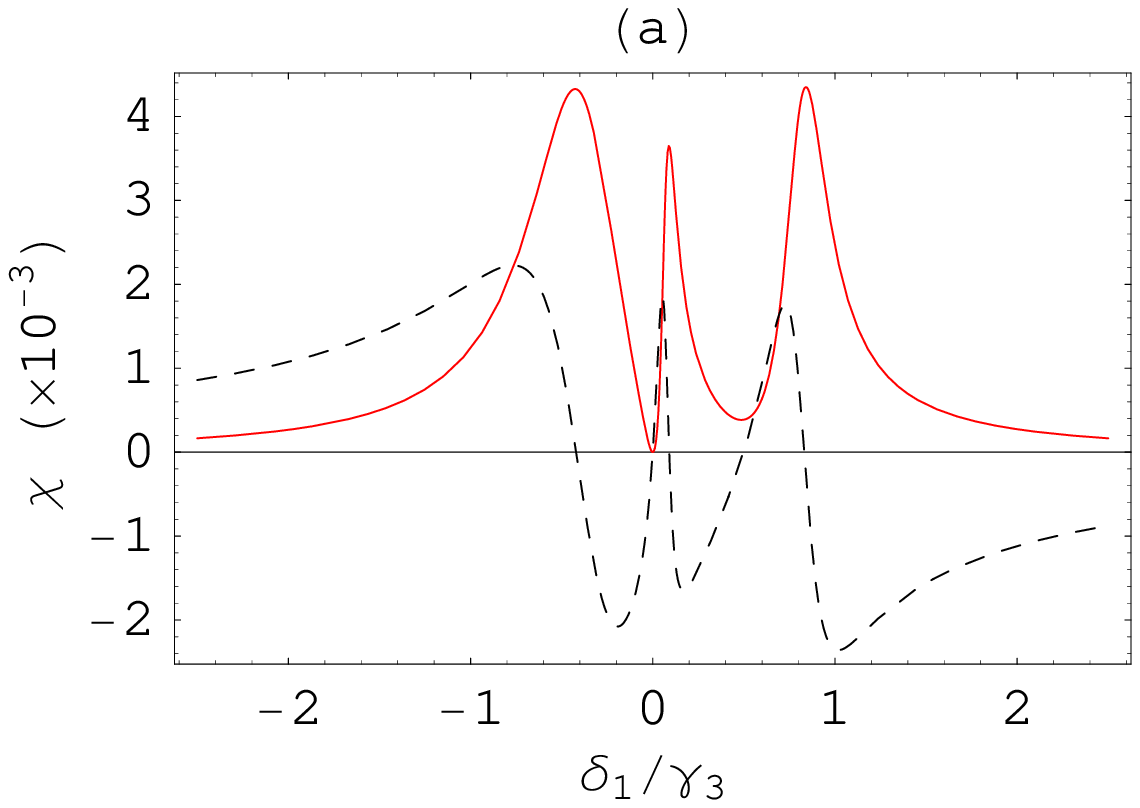}
    \includegraphics[width=8cm]{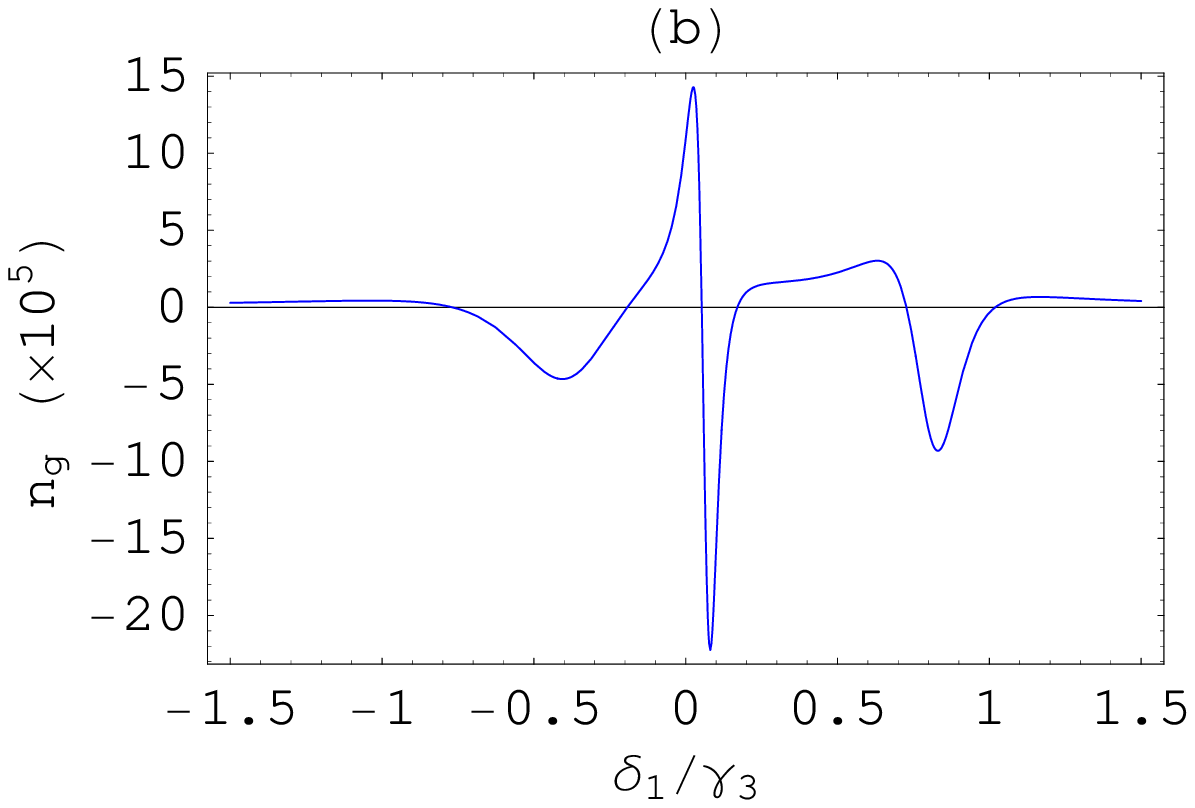}
    \includegraphics[width=8cm]{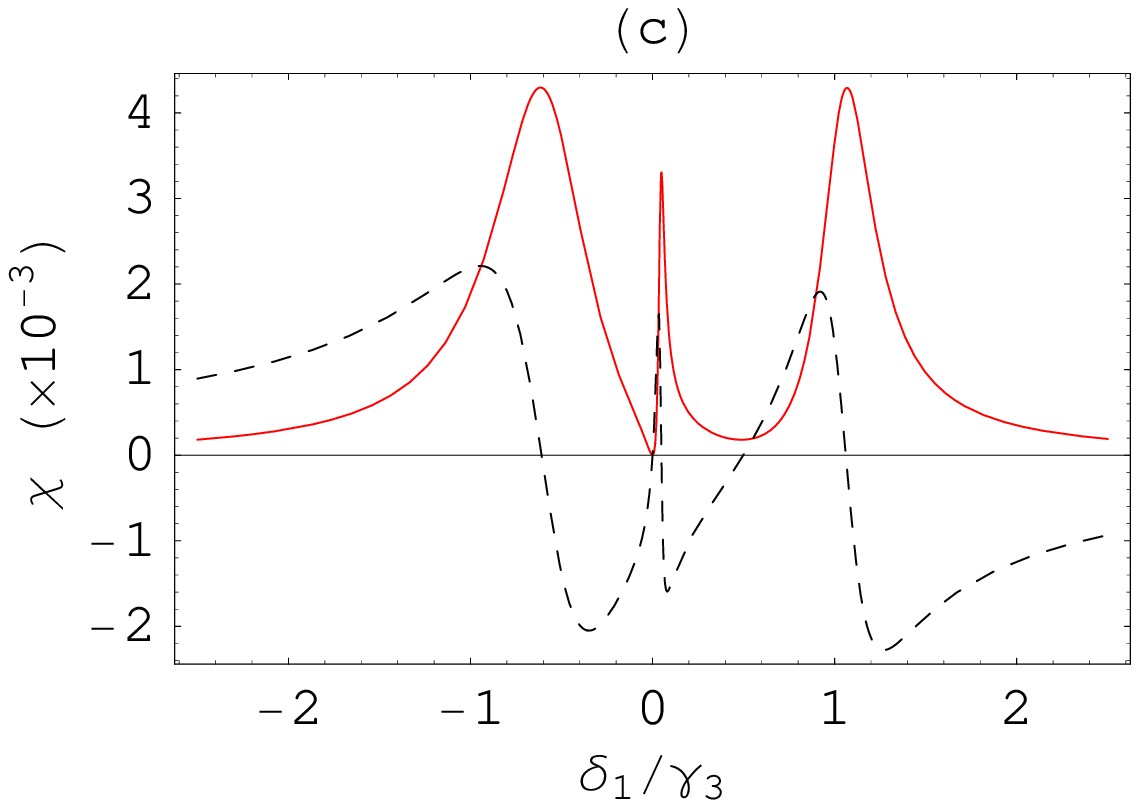}
    \includegraphics[width=8cm]{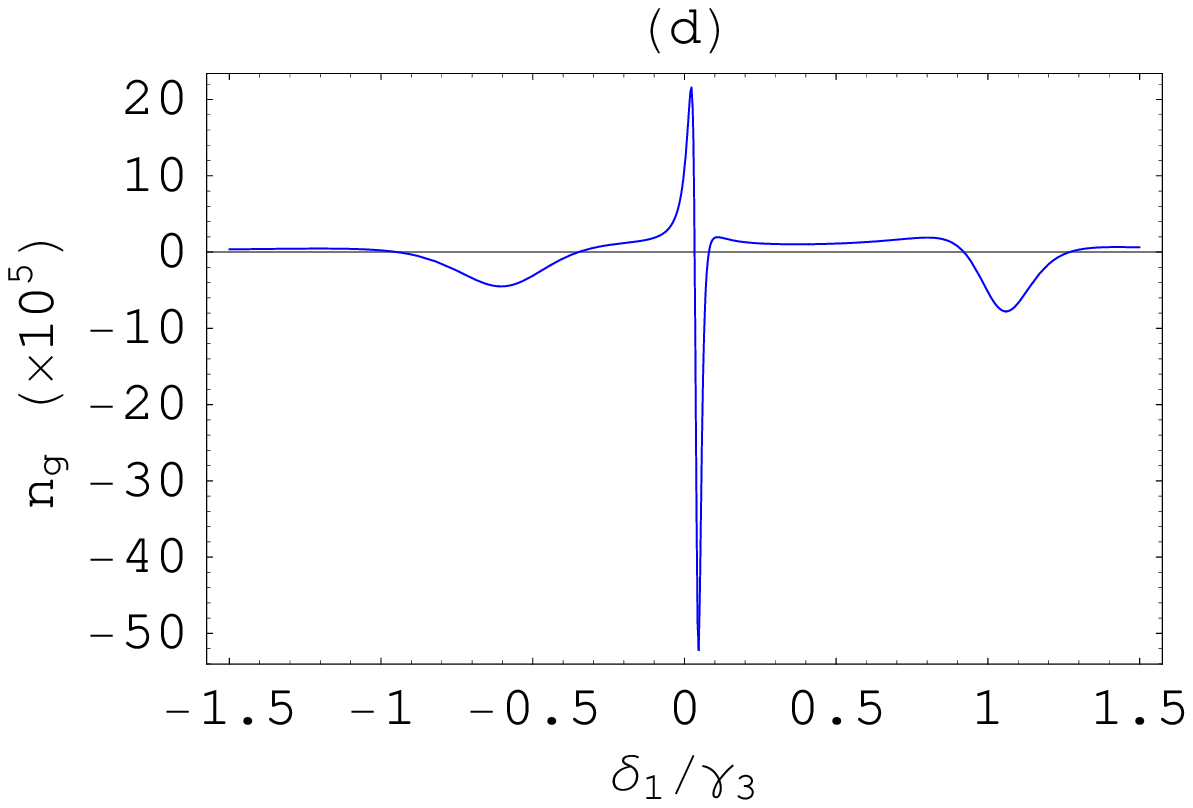}
    \includegraphics[width=8cm]{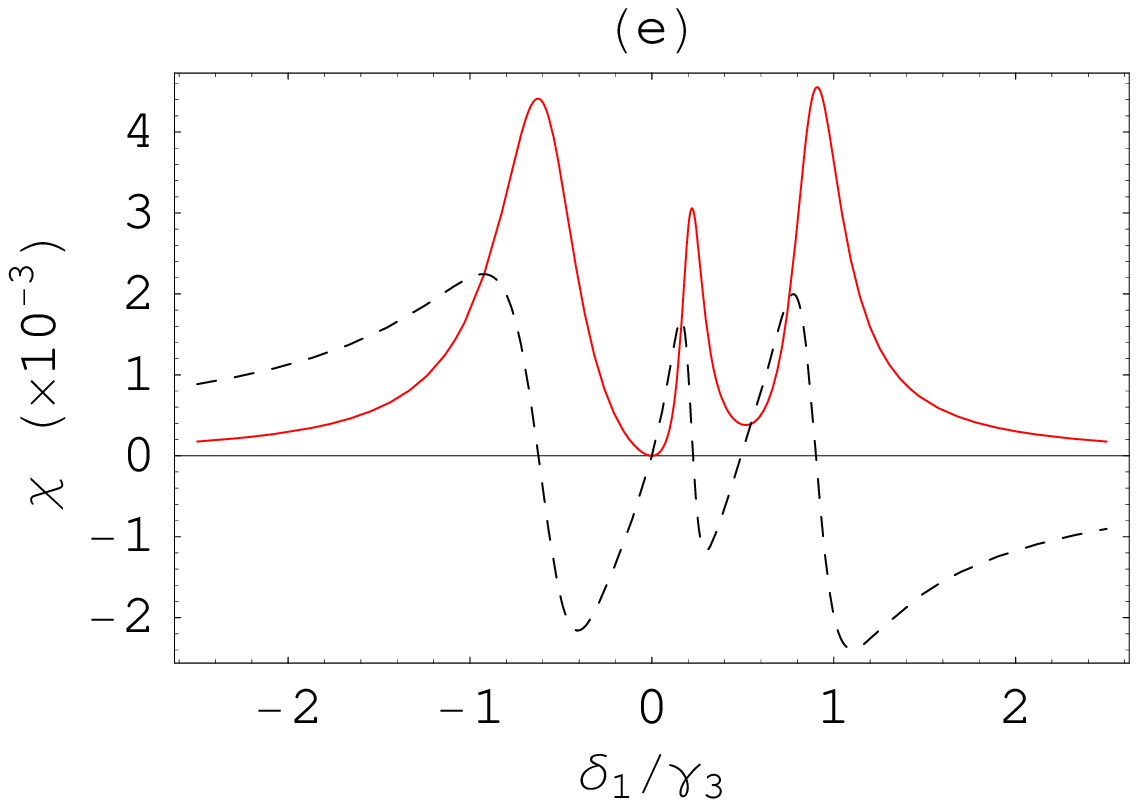}
    \includegraphics[width=8cm]{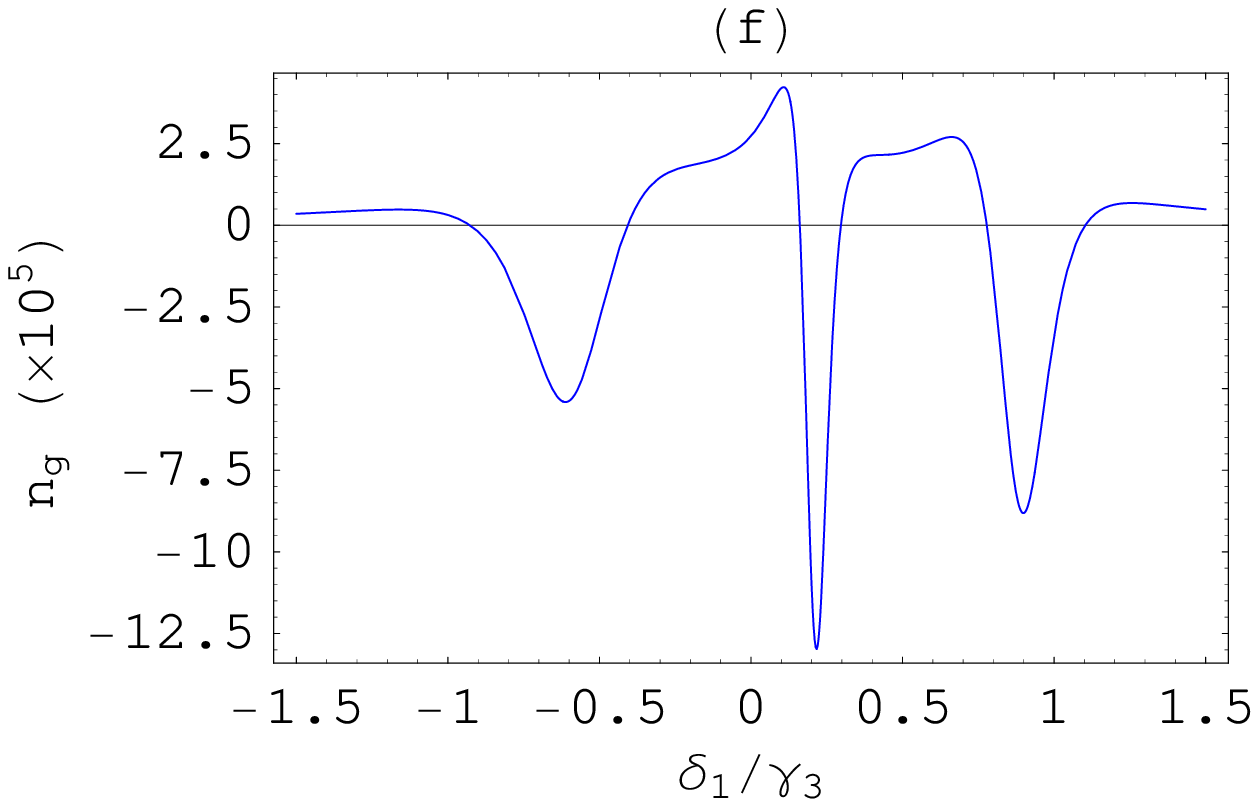}
    \includegraphics[width=8cm]{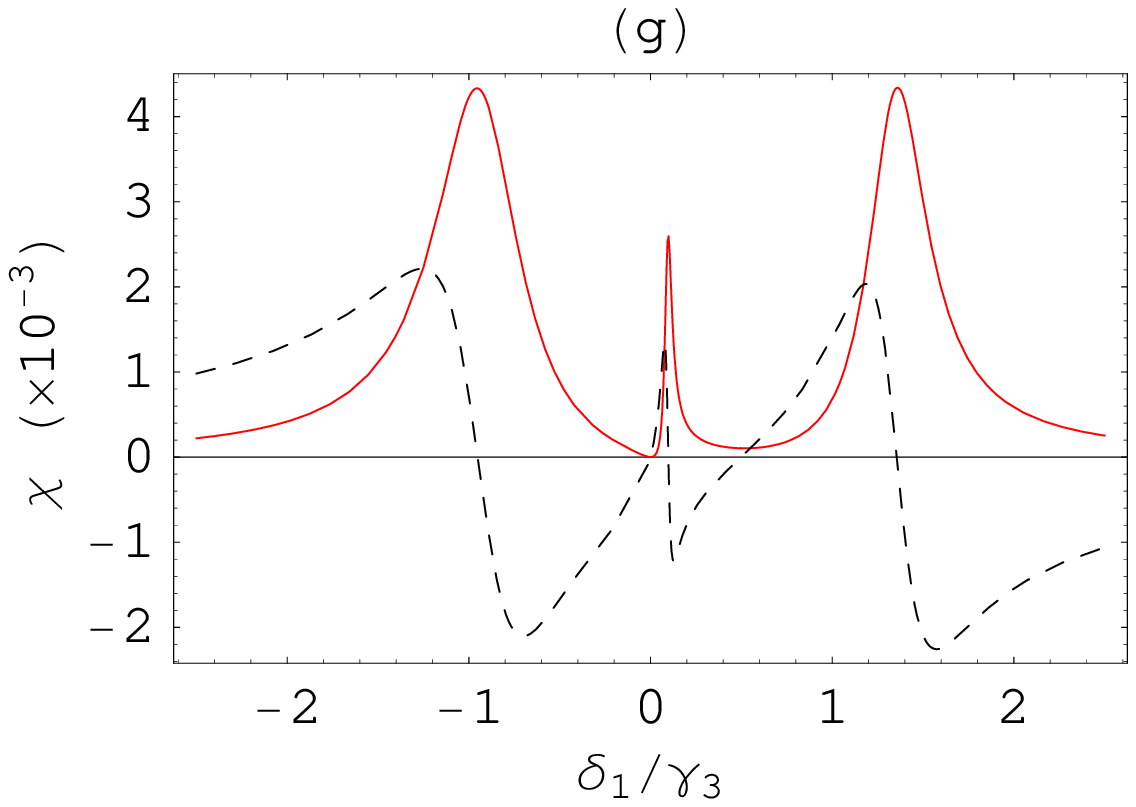}
    \includegraphics[width=8cm]{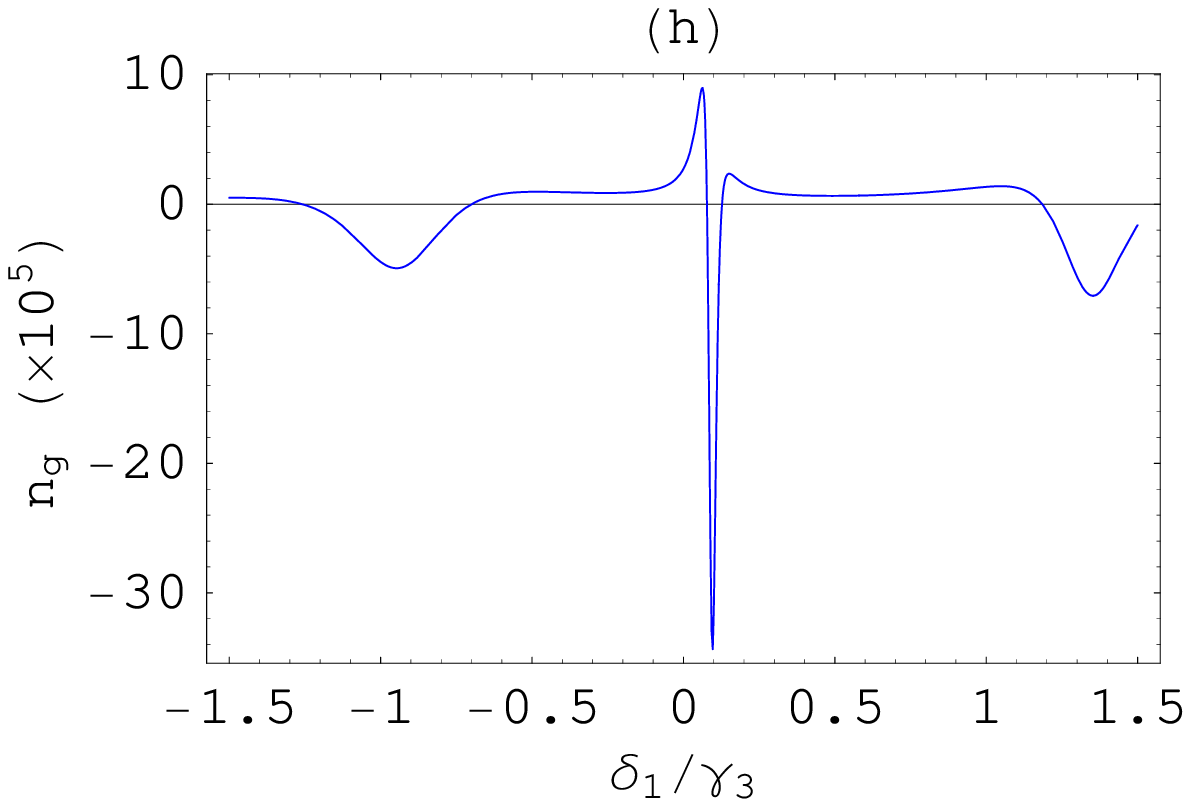}
\caption{(Color online) Left panel: Absorption (solid line) and dispersion (dashed line) of the
probe laser (L1); Right panel: the corresponding group index. The parameters for the calculations are:
(a)-(b) $\Omega_3=1.0\gamma_3$, $\Omega_2=0.5\gamma_3$.
(c)-(d) $\Omega_3=1.5\gamma_3$, $\Omega_2=0.5\gamma_3$.
(e)-(f) $\Omega_3=1.0\gamma_3$, $\Omega_2=1.0\gamma_3$.
(g)-(h) $\Omega_3=2.0\gamma_3$, $\Omega_2=1.0\gamma_3$.
Other parameters are: $\delta_2=0$, $\delta_3=-0.5\gamma_3$, $\gamma_4=0.1\gamma_3$.}
\end{figure}
\subsection{The Group Velocity}
It is clear that the absorption spectrum of the probe laser depends on both the detuning and the Rabi frequency of the coupling lasers. We plot the imaginary part(absorption) and the real part(dispersion) of the linear susceptibility of the probe laser in the same frame as a function of the frequency detuning of the probe laser in figure 3(a). The dispersion curves(see dashed line) display a very high positive slope at the center of the transparency windows; therefore the group velocity of the probe laser can be slowed down without absorption.
The group velocity of the probe field can be calculated by $v_{g}=c/n_{g}$, where c is the speed of light in vacuum  and the group velocity index is given by~\cite{PRA.66.015802.2002}
\begin{eqnarray}
n_{g}=1+\frac{1}{2}\chi^{'}+\frac{\nu_1}{2}\frac{\partial\chi^{'}}{\partial\nu_{1}},
\end{eqnarray}
which is evaluated at the carrier frequency of the probe laser.

The group index $n_g$ based on equation (27) is simultaneously plotted in figure 3(b) as a function of the frequency detuning of the probe laser corresponding to the same parameters as in figure 3(a). One can see that the group velocity of the probe laser can be reduced by as much as a factor of $10^6$ for the chosen parameters without absorption. When the Rabi frequency of laser L3 increases as shown in figure 3(c), the absorption corresponding to the transparency window at $\delta_1=-\delta_3$ decreases and the width of the transparency window increases, while the slope of the dispersion curves decreases at $\delta_1=-\delta_3$, and therefore the group index at $\delta_1=-\delta_3$ decreases as shown in figure 3(d). When the Rabi frequency of the laser L2 increases the width of the EIT window at $\delta_1=\delta_2$ increases and the central absorption peak is pushed toward to the second EIT window at $\delta_1=-\delta_3$ as shown in figure 3(e), while the group index decreases at $\delta_1=\delta_2$, and increases at $\delta_1=-\delta_3$ as shown in figure 3(f). When we further increase $\Omega_3$ the central absorption component is pushed back toward the EIT window at $\delta_1=\delta_2$ and the second transparency window becomes wider and deeper as shown in figure 3(g). The slopes of the dispersion curves as well as the group index at the EIT windows decreases accordingly as shown in figure 3(h). Clearly, the absorption at the two EIT windows, and therefore the corresponding group index can be very different depending on the coupling lasers. Consequently, the EIT can be controlled by the detunings and Rabi frequencies of two coupling lasers as well as the group velocities at these two windows. This can be very useful for quantum information processing and transfer. One can control the propagation of probe signals at two adjacent frequencies with the two coupling fields. To uncover the responsible physical parameters for the group index within each EIT window, we write the group index equation (27) explicitly.

(1) For the $\delta_1=\delta_2$ transparency window:
\begin{eqnarray}
n_g(\delta_1=\delta_2)=1+2\kappa\times\left(\frac{\delta_2+\omega_{31}}{\Omega_2^2}\right),
\end{eqnarray}
with $\kappa=\frac{N|\mu_{13}|^2}{\epsilon_{0}\hbar}$, which is a function of $\delta_2$ and $\Omega_2$ of laser L2, but independent of laser L3. For a given detuning $\delta_2$, the group index at the $\delta_1=\delta_2$ transparency window is inversely proportional to the intensity of laser L2 as illustrated in figure 4(a).

(2) For the $\delta_1=-\delta_3$ transparency window:
\begin{eqnarray}
n_g(\delta_1=-\delta_3)=1-\frac{\kappa(\delta_2+\delta_3)\beta\gamma_4}
                                {4(\alpha^2+\beta^2)}\nonumber\\
                                +\frac{\kappa(\frac{\omega_{31}-\delta_3}{2})
                                    \left(\beta\frac{\gamma_4}{2}-(\delta_2+\delta_3)
                                    (\alpha+\frac{\partial\beta}{\partial\delta_1}
                                     \frac{\gamma_4}{2})\right)}{\alpha^2+\beta^2}\nonumber\\
                                     +\frac{\kappa\gamma_4(\frac{\omega_{31}-\delta_3}{4})
                                     (\delta_2+\delta_3)\beta
                                     (\alpha\frac{\partial\alpha}{\partial\delta_1}
                                     +\beta\frac{\partial\beta}{\partial\delta_1})}
                                {(\alpha^2+\beta^2)^2}
\end{eqnarray}
with
\begin{eqnarray}
\alpha=\frac{(\delta_3+\delta_2)(\omega_3^2+\gamma_3\gamma_4/4)}{4}\nonumber\\
\beta=\frac{\gamma_4}{2}\left(\Omega_2^2/4-\delta_3(\delta_2+\delta_3)\right)\nonumber\\
\frac{\partial\alpha}{\partial\delta_1}=\frac{\Omega_2^2+\Omega_3^2+\gamma_3\gamma_4/4}{4}
                                    -\delta_3(\delta_3+3\delta_2)\nonumber\\
\frac{\partial\beta}{\partial\delta_1}=-\delta_3(\gamma4+\gamma_3/2)-\delta_2\frac{\gamma_3+\gamma_4}{2}\nonumber
\end{eqnarray}
\begin{figure}
    \includegraphics[width=3 in]{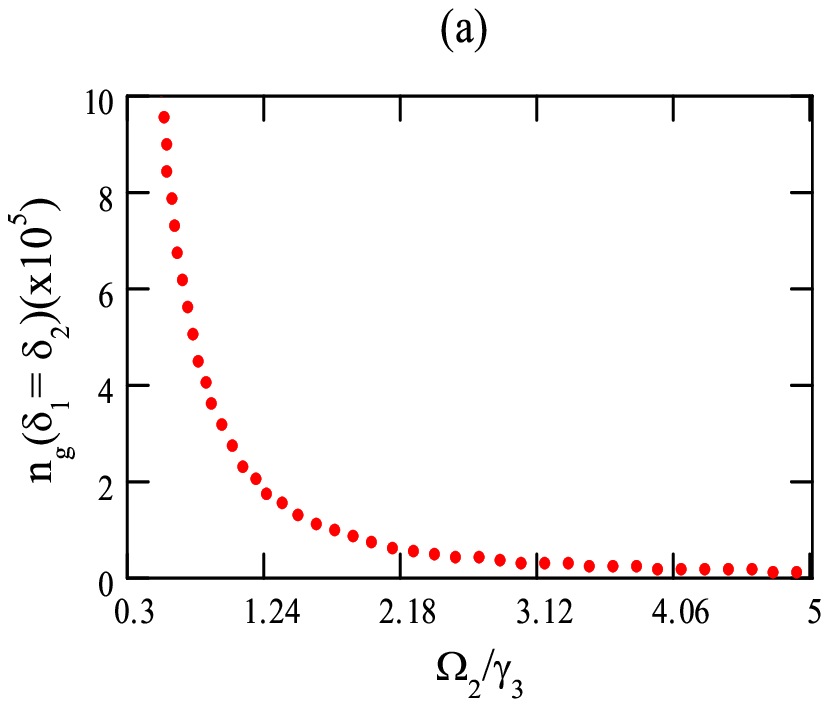}
    \includegraphics[width=3 in]{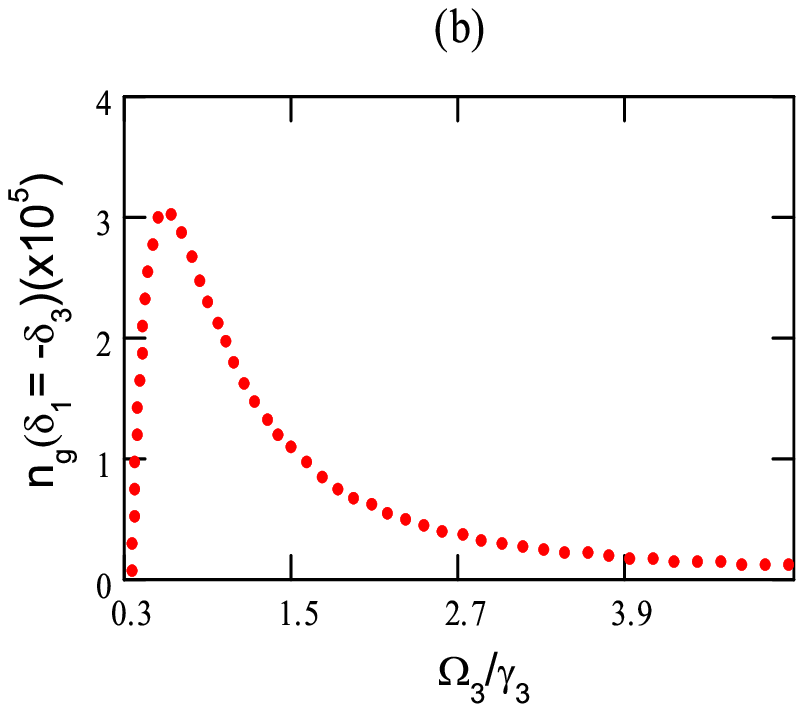}
    \includegraphics[width=3 in]{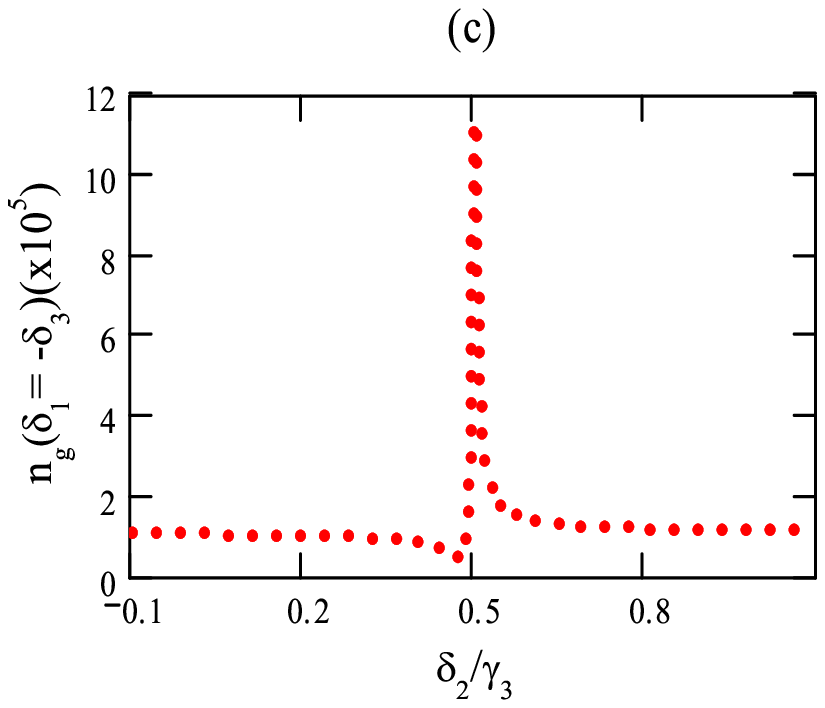}
    \includegraphics[width=3 in]{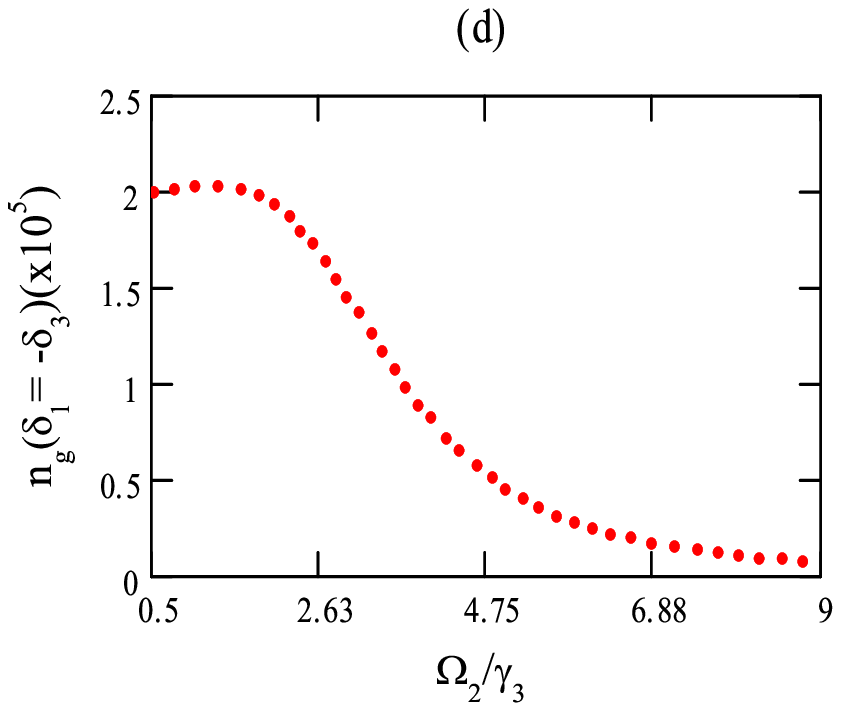}
\caption{(Color online) (a) The group index at the center of the EIT window,  $\delta_1=\delta_2$, as a function of $\Omega_2$. Other parameters are $\delta_2=0.3\gamma_3$, $\delta_3=0$, $\Omega_3=1.5\gamma_3$ and $\gamma_4=0.1\gamma_3$. The group index at the center of the EIT window $\delta_1=-\delta_3=-0.5\gamma_3$: (b) as a function of $\Omega_3$ for $\delta_2=0$ and $\Omega_{2}=0.5\gamma_3$; (c) as a function of $\delta_2$ for $\Omega_{3}=1.5\gamma_3$ and $\Omega_{2}=0.5\gamma_3$; (d) as a function of $\Omega_2$, for $\delta_2=0$ and $\Omega_{3}=1.0\gamma_3$.}
\end{figure}
However, at the transparency window $\delta_1=-\delta_3$, the group index depends on both lasers and it is a function of the three parameters $\delta_2$, $\Omega_2$ and $\Omega_3$ for a given $\delta_3$. In figure 4(b)-(d) we show the group index as a function of $\Omega_3$, $\delta_2$, or $\Omega_2$, respectively, when the two other parameters are given at the transparency window $\delta_1=-\delta_3$. This clearly shows that the group index can be manipulated by the parameters of both coupling lasers.

When both coupling lasers are on resonance or in Raman detuning with each other($\delta_2=-\delta_3$), the two transparency windows merge into one as shown in figure 5(a) and 5(b), respectively. The width of the transparency window increases with the Rabi frequencies, $\Omega_2$ and $\Omega_3$, as shown in figure 5(c). If we inspect the spectrum carefully we notice that there is a small feature at the center of the transparency window. We plot it in an expanded scale in figure 5(d). We find that the dispersion changes between normal and abnormal within a very narrow frequency region. Consequently, the group index changes from negative to positive; in other words, the probe laser can be switched from a anomalous dispersion associated with a negative group index(or superluminal) to a positive group index(or subluminal) without being absorbed within this window. Although the transparency is not one hundred percent for the negative group index region the absorption is small.
\begin{figure}
    \includegraphics[width=8cm]{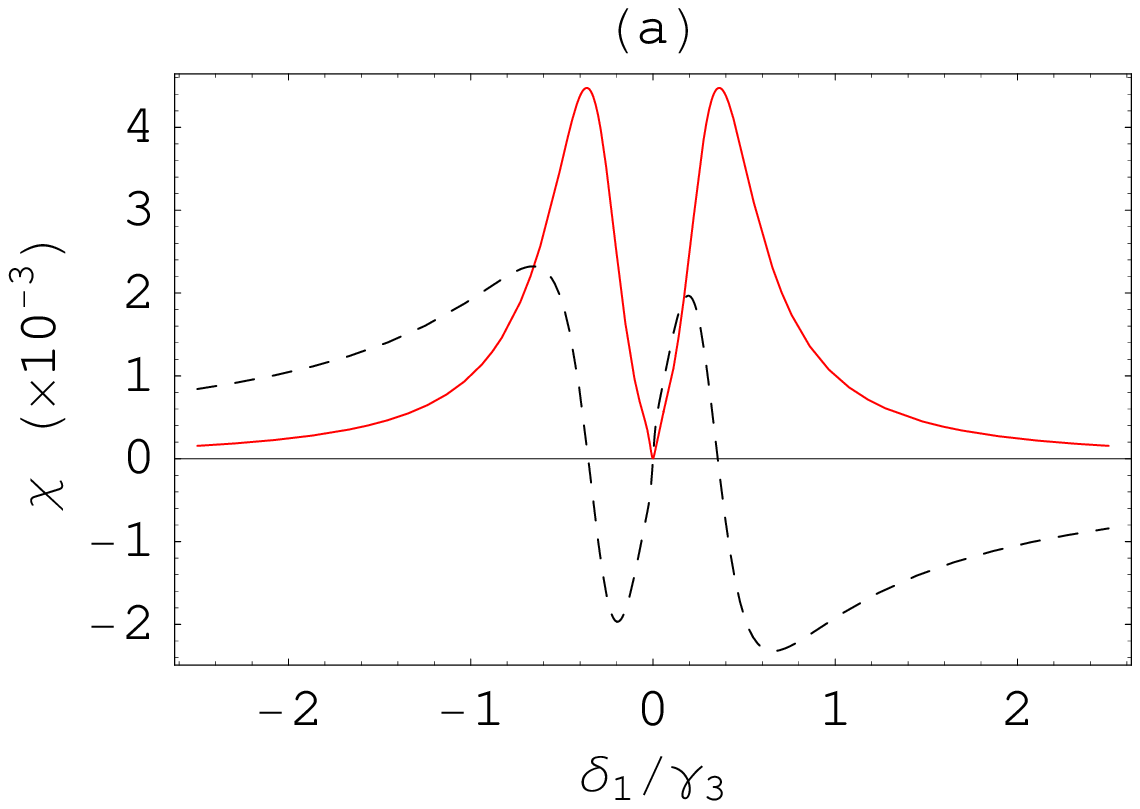}
    \includegraphics[width=8cm]{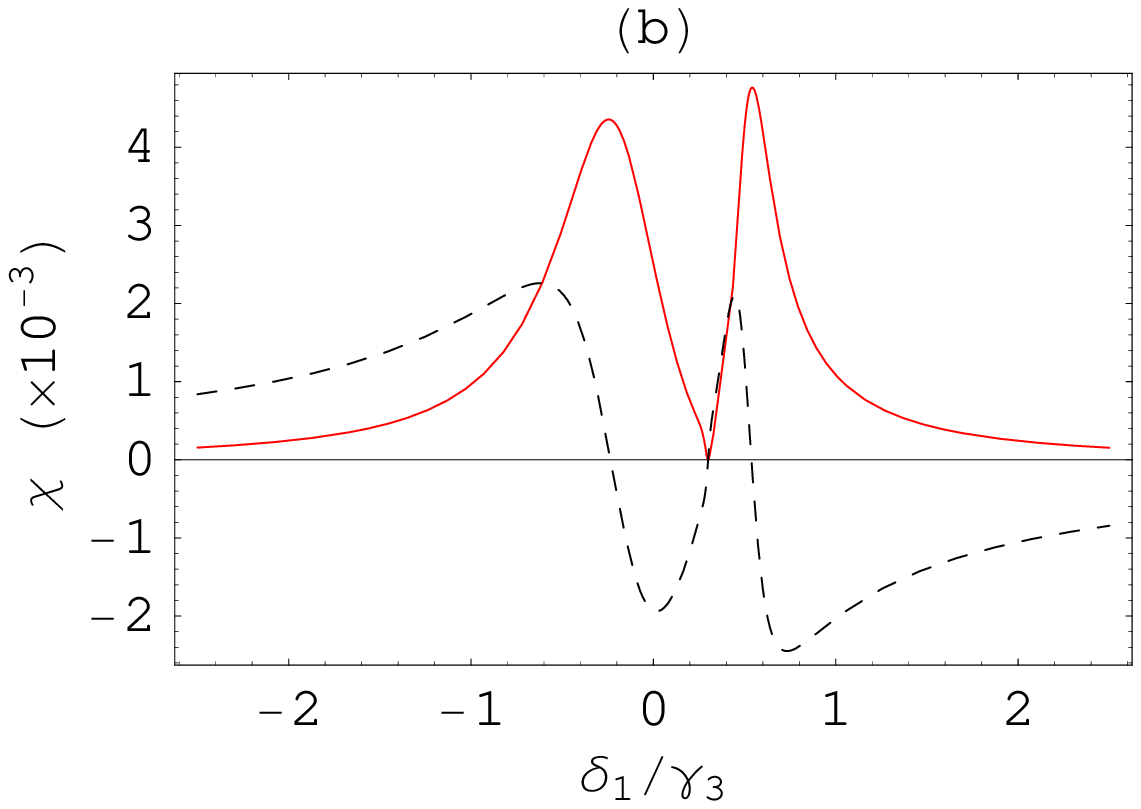}
    \includegraphics[width=8cm]{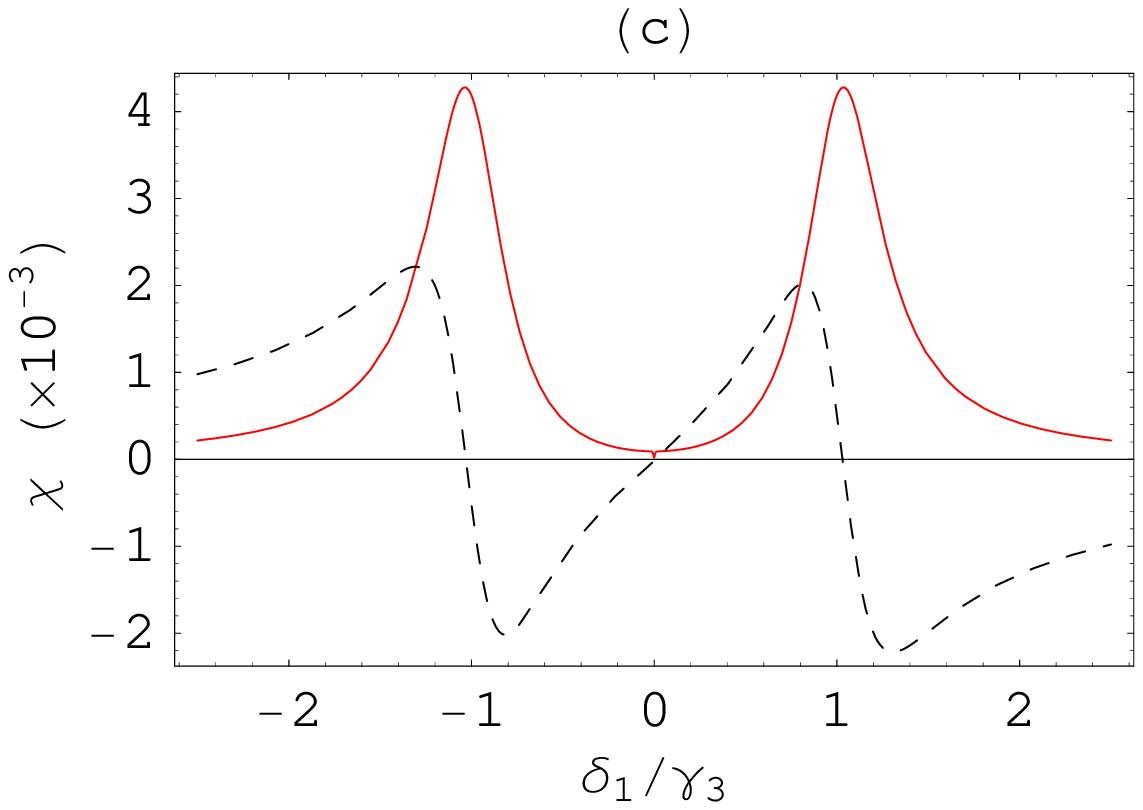}
    \includegraphics[width=8cm]{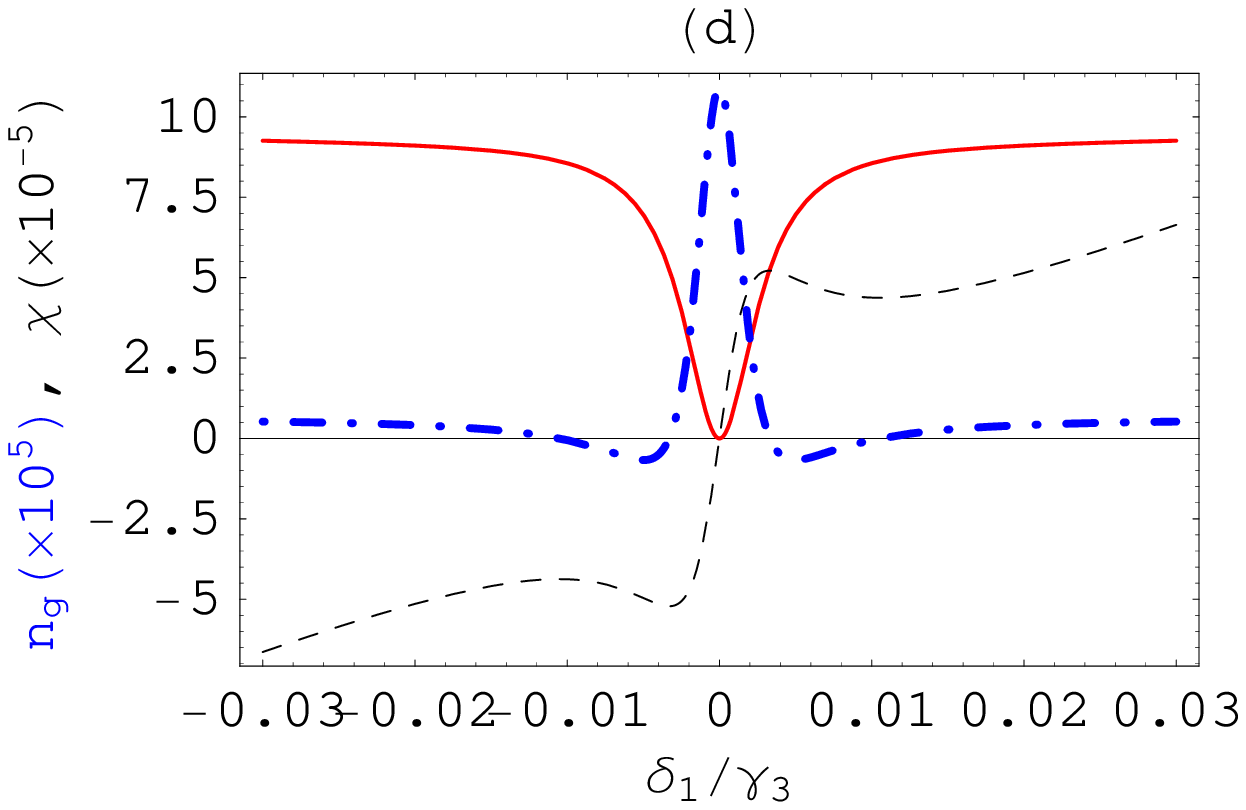}
\caption{(Color online) The susceptibility and the group index as a function of the probe detuning $\delta_1$ when two coupling lasers are on resonance or on Raman detuning($\delta_2=-\delta_3$). (a) $\delta_2=\delta_3=0$, $\Omega_{3}=\Omega_{2}=0.5\gamma_3$. (b) For laser L2 and L3 are on Raman detuning: $\delta_2=-\delta_3=0.3\gamma_3$, $\Omega_{3}=\Omega_{2}=0.5\gamma_3$. (c) $\delta_2=\delta_3=0$, $\Omega_{3}=2.0\gamma_3$, and $\Omega_{2}=0.5\gamma_3$. (d) we plot (c) on an expanded scale to show the details within the EIT window(solid lines: imaginary part of $\chi$, dashed lines: the real part of $\chi$; dotdashed lines: the group index).}
\end{figure}

Based on the above analysis, the scheme can be realized in a four-level rubidium $^{87}Rb$ atom. Two hyperfine ground state levels, $|1\rangle=|5S_{1/2}$, $F''=1\rangle$, and $|2\rangle=|5S_{1/2}$, $F''= 2\rangle$ are coupled by laser L1 and laser L2 to a common intermediate excited hyperfine level, $|3\rangle=|5P_{1/2}$, $F'=1\rangle$, respectively. The third laser L3 couples the intermediate level to a higher excited $5D_{3/2}$ hyperfine level $|3\rangle=|5D_{3/2}$,$F =2\rangle$. In order to observe the phenomena experimentally, an ultracold $^{87}Rb$ atomic ensemble formed in a MOT with a typical temperature of 100 $\mu{K}$ would satisfy all the conditions corresponding to our above calculations. The lifetime of $5D_{3/2}$ state is very long compared to that of the $5P_{1/2}$ state($\gamma_4\simeq0.1\gamma_3$)~\cite{http.steck.us.alklidata}. So all the conditions assumed in our analysis can be satisfied in this system. We expect the experimental observations can be readily realized.
\section{Summary}
In conclusion, we have shown that the response of a probe laser in an inverted Y-type four-level system driven by two additional coherent fields exhibits double transparency windows for the probe laser. The reliability of the calculations is established by the agreement in the susceptibility of the probe laser obtained by both wavefunction and density matrix methods. The transparency windows can be controlled by the amplitude and frequency detuning of the coupling fields. The group index associated with the group velocity of the probe laser can be very different at the two transparency windows; hence it can be controlled by the coupling fields. The propagation of the probe field can be switched from superluminal near the resonance to subluminal on resonance of the probe transition within the single transparency window when the two coupling lasers are detuned on resonance. This provides a potential application in quantum information processing. This scheme may be realized in an ultracold $^{87}Rb$ system and can be used to investigate both superluminal and slow light.
\section{Acknowledgement}
This work is supported by the Research Development Grant from Penn State University.

\end{document}